\documentclass[twocolumn,pra,aps,showpacs,superscriptaddress]{revtex4-2}
\usepackage[english]{babel}
\usepackage[utf8]{inputenc}

\usepackage{url,psfrag,graphicx}
\usepackage{dcolumn}
\usepackage{amsmath,amssymb,amsthm}
\usepackage{bm}
\usepackage{pstricks}
\usepackage{hyperref}
\usepackage{epsfig,placeins}
\usepackage{mathtools}
\usepackage{commath}
\usepackage{subfigure}
\usepackage{multirow}

\def\pl{\partial}

\def\<{\left<}
\def\>{\right>}
\def\ket|#1>{\left|#1\right>}
\def\bra<#1|{\left<#1\right|}
\def\elem<#1|#2|#3>{\left<#1\right|#2\left|#3\right>}
\def\nn{\nonumber}

\def\HH{\text{H}_2}

\def\Tr{\text{\rm Tr}}

\def\({\left(}
\def\){\right)}

\def\beq{\begin{equation}}
\def\eeq{\end{equation}}

\begin{document}

\title{Long term behavior of the stirred vacuum on a Dirac chain:
  geometry blur and the random Slater ensemble}

\author{José Vinaixa}
\affiliation{Dto. Física Fundamental, Universidad Nacional de
  Educación a Distancia (UNED), Madrid, Spain}

\author{Begoña Mula}
\affiliation{Dto. Física Fundamental, Universidad Nacional de
  Educación a Distancia (UNED), Madrid, Spain}

\author{Alfredo Deaño}
\affiliation{Dto. Matemáticas, Universidad Carlos III de Madrid,
  Leganés, Spain}

\author{Silvia N. Santalla}
\affiliation{Dto. Física \&\ GISC, Universidad Carlos III de Madrid,
  Leganés, Spain}

\author{Javier Rodríguez-Laguna}
\affiliation{Dto. Física Fundamental, Universidad Nacional de
  Educación a Distancia (UNED), Madrid, Spain}

\begin{abstract}
  We characterize the long-term state of the 1D Dirac vacuum stirred
  by an impenetrable object, modeled as the ground state of a finite
  free-fermionic chain dynamically perturbed by a moving classical
  obstacle which suppresses the local hopping amplitudes. We find two
  different regimes, depending on the velocity of the obstacle. For a
  slow motion, the effective Floquet Hamiltonian presents features
  which are typical of the Gaussian orthogonal ensemble, and the
  occupation of the Floquet modes becomes roughly homogeneous.
  Moreover, the long term entanglement entropy of a contiguous block
  follows a Gaussian analogue of Page's law, i.e. a volumetric
  behavior. Indeed, the statistical properties of the reduced density
  matrices correspond to those of a random Slater determinant, which
  can be described using the Jacobi ensemble from random matrix
  theory. On the other hand, if the obstacle moves fast enough, the
  effective Floquet Hamiltonian presents a Poissonian behavior. The
  nature of the transition is clarified by the entanglement links,
  which determine the effective geometry underlying the entanglement
  structure, showing that the one-dimensionality of the physical
  Hamiltonian dissolves into a random adjacency matrix as we slow down
  the obstacle motion.
\end{abstract}

\date{October 24, 2023}

\maketitle


\section{Introduction}
\label{sec:intro}

One of the most relevant insights obtained from quantum mechanics is
the fact that a static {\em vacuum} is merely the ground state (GS) of
a certain Hamiltonian. Therefore, its structure can be quite complex,
and may present very relevant physical effects. For example, when a
piece of vacuum is constrained by movable walls, they can feel Casimir
forces \cite{Casimir.48}. If these walls move, they can induce
transitions to excited states \cite{Moore.70, Alves.03}.
Interestingly, the vacuum state typically presents quantum
correlations, leading to entanglement between different regions.
Moreover, the relation between the vacuum entanglement and geometry is
known to run deeper than expected. For example, many low-energy
quantum states respect the {\em area law}, i.e. the entanglement
entropy (EE) between a region and its environment is proportional to
the measure of its boundary \cite{Srednicki.93, Eisert.10}. The area
law has been rigorously proved in a few cases, such as the GS of
gapped 1D Hamiltonians \cite{Hastings.07}. Yet, it receives
logarithmic corrections in many critical states, as it is predicted by
conformal field theory (CFT) \cite{Calabrese.04, Calabrese.09}. In
fact, it is possible to read the underlying geometry of a quantum
state without knowledge of the associated Hamiltonian, making use of
the so-called {\em entanglement link} (EL) representation
\cite{Singha.20, Singha.21, Santalla.23}.

In this article we describe the long-term behavior of a portion of the
1D Dirac vacuum stirred by an impenetrable object moving through it.
As a mental image, we may think of a piston moving through an empty
cylinder, which would have no classical effect, but will have a
considerable effect in quantum mechanics \cite{Stefanatos.13}. To that
end, we define a toy model, which we call the {\em stirred Dirac
  vacuum}. In it, we start out with the GS of a free-fermionic chain,
which can be used as a model of the Dirac vacuum in (1+1)D, and can be
implemented physically using ultracold atoms in optical lattices
\cite{Jaksch.05, Lewenstein.12, Gross.17}. We then introduce a classical
obstacle, which acts like a movable boundary condition, canceling the
local hopping amplitudes. This obstacle is forced to move forward,
thus injecting energy into the system, and repeating its motion after
reaching the end. In the long run, the instantaneous physical states
will define a certain ensemble, which will depend on the velocity at
which the obstacle moves. This dependence on the velocity of the
quench has been highlighted in a variety of situations
\cite{Peschel.11, Goldman.14, Bukov.16, Seetharam.17}.

Since our system is subject to a periodic perturbation, we may
describe it using a Floquet effective Hamiltonian \cite{Eckardt.17,
  DAlessio.14, Lazarides.14, Ponte.15, Moessner.17, Tsuji.23}, whose
long term behavior may be described using random matrix theory (RMT)
\cite{Berry.77, Bohigas.84, Mehta.04, Scaramazza.16}. Moreover, random
quantum states chosen according to a unitary-invariant measure are
known to present {\em volumetric entanglement} and follow Page's law
\cite{Page.93, Bianchi.22}. Recently, the analogue for Gaussian states
has been described \cite{Magan.16, Bianchi.21, Huang.22}, based on
previous results from random matrix theory \cite{Zyczkowski.99,
  Collins.04}, allowing us to characterize a {\em random Slater
  ensemble} (RSE). Yet, the approach of our stirred Dirac vacuum
towards the RSE can be hindered by the existence of preserved
quantities. As we will show, the values of the occupations of the
Floquet modes within the initial state allow us to predict whether the
RSE will be finally achieved or not, thus characterizing two different
regimes, which can be further distinguished through their entanglement
geometry via the aforementioned EL representation \cite{Singha.20,
  Singha.21, Santalla.23}, showing that in the slow stirring phase the
initial geometry is effectively blurred, while it remains if the
stirring is fast.

This article is organized as follows. Sec. \ref{sec:intro} describes
our model and the simulation procedure. Then we describe our first
numerical results in Sec. \ref{sec:num}, based on the energy
absorption and the Floquet effective Hamiltonian, finishing with an
analysis of the Floquet occupations. In Sec. \ref{sec:rmt} we provide
the necessary background regarding the RSE, and we apply it to
describe the long-term behavior of the slow phase in Sec.
\ref{sec:entang}. The transition between the slow and the fast phases
is described in Sec. \ref{sec:links} using the entanglement links. The
article finishes summarizing our conclusions and proposals for further
work.


\section{The stirred Dirac vacuum}
\label{sec:intro}

\begin{figure}
  \includegraphics[width=7cm]{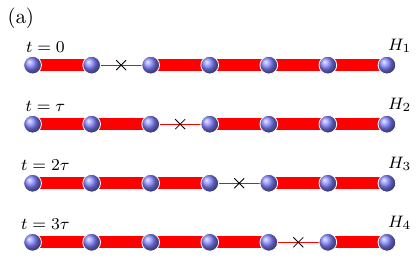}
  \vskip 3mm
  \includegraphics[width=7cm]{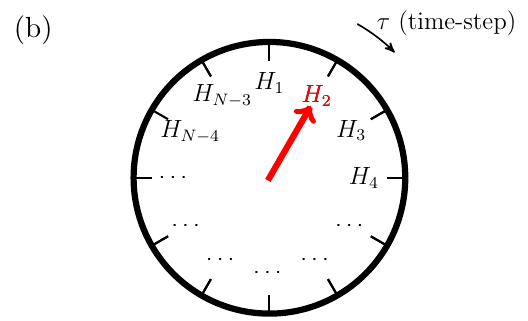}
  \caption{(a) Illustration of the discrete obstacle dynamics. A
    particular sequence of Hamiltonians $\{H_i\}$ is used to create a
    time-dependent $H(t)$, switching each $H_i$ after a time-step
    $\tau$. It may described as an impenetrable obstacle moving along
    the chain. (b) This sequence may be represented as being driven by
    a stepper motor adjusted to a time delay $\tau$, with total
    periodicity $T=(N-3)\tau$.}
  \label{fig:illust}
\end{figure}

Let us build a discrete analogue of the 1D Dirac vacuum on an $N$-site
free-fermionic chain. In order to do that, we define {\em link}
operators,

\beq
L_i \equiv -{1\over 2} \(c^\dagger_i c_{i+1} + \text{h.c.}\),
\eeq
where $c^\dagger_i$ and $c_i$ are (spinless) fermionic creation and
annihilation operators, with $i\in \{1\cdots N\}$, and the 1/2 factor
has been chosen for later convenience. Now, let us define our initial
Hamiltonian, on a chain with open boundaries,

\beq
H_0\equiv \sum_{i=1}^{N-1} L_i,
\label{eq:h0}
\eeq
whose ground state (GS), containing $m=N/2$ particles, will constitute
the initial state for our simulations, $\ket|\Psi(0)>$. Now, let us
define a family of Hamiltonians,

\beq
H_i\equiv H_0 - L_{i+1},
\eeq
such that $H_i$ has a broken link between sites $i+1$ and $i+2$, as it
is depicted in Fig. \ref{fig:illust} (a). Our system will be subject
to Schrödinger's equation

\beq
i\pl_t\ket|\Psi(t)>= H(t)\ket|\Psi(t)>,
\eeq
where

\beq
H(t)=H_i, \qquad \text{if $t\in[(i-1)\tau,i\tau]$},
\eeq
i.e.: $H(t)$ equals $H_1$ if $t\in [0,\tau]$, $H_2$ if $t\in
[\tau,2\tau]$, etc., up to time $T=(N-3)\tau$, check Fig.
\ref{fig:illust} (b). After that time, the full sequence repeats,

\beq
H(t+T)=H(t).
\eeq

\bigskip

Therefore, at each instant $t$, one link of the chain will be absent,
effectively splitting the initial chain into two disconnected parts,
as if an impenetrable obstacle was interposed. The broken link will
move along the chain, always rightwards, spending a fixed time $\tau$
on each position, and never leaving an isolated site at any time, and
repeating its full pattern after time $T=(N-3)\tau$. Also notice that
the obstacle moves in discrete steps, with an average velocity
$v=1/\tau$. Thus, the only relevant parameters of our model are the
size $N$ and the time-step $\tau$. As we will see, {\em slow
  schedules} $\tau\gg 1$ and {\em fast schedules} $\tau\ll 1$ lead to
a very different long-term behaviors of the system.

Let us stress that, in classical terms, the movement of the obstacle
through the vacuum should bear no effect on its properties. Yet, the
quantum nature of the state gives rise to an amplification of the
vacuum fluctuations, and pairs particle-antiparticle will appear, in
similarity to the dynamical Casimir effect \cite{Moore.70,Alves.03}.

\bigskip

In the remainder of this section we will review some basic properties
of the time-evolved quantum vacuum within our model.

The instantaneous Hamiltonian $H(t)$ can always be diagonalized in
single-body terms. Therefore, the state $\ket|\Psi(t)>$ can always be
written as a Gaussian state, or Slater determinant, which we can write
as

\beq
\ket|\Psi(t)>=\prod_{k=1}^{N/2} b^\dagger_k(t) \ket|0>,
\label{eq:slater}
\eeq
where $\ket|0>$ is the Fock vacuum, and the creation operators
$b^\dagger_k(t)$, which are usually called the occupied modes, can be
written as

\beq
b^\dagger_k(t)=\sum_{i=1}^N U_{ki}(t)\; c^\dagger_i,
\eeq
and $U_{ki}(t)$ are entries of an $N\times N$ unitary matrix, $U(t)$,
which satisfies

\beq
i\pl_t U(t)=H(t) U(t).
\eeq

The instantaneous correlation matrix is defined as

\beq
C_{ij}(t)\equiv \bra<\Psi(t)|c^\dagger_i c_j\ket|\Psi(t)>= \sum_k \bar
U_{ki}(t) U_{kj}(t),
\eeq
and can be considered as a projector on the set of occupied modes,
since it is hermitian and its spectrum is contained in $\{0,1\}$.
Moreover, let us stress that the single-body Hamiltonian, i.e. the
matrix $h_{ij}(t)$ such that

\beq
H(t)=-\sum_{ij} h_{ij}(t) c^\dagger_i c_j,
\eeq
is {\em bipartite}, i.e. there exists a bipartition of the set of
sites (even vs.~odd in our case) such that $h_{ij}(t)$ is nonzero only if
$i$ and $j$ belong to opposite parts. This implies that the density in
the GS,

\beq
\<n_i\>=\langle c^\dagger_i c_i\rangle=1/2,
\eeq
for all times and that the instantaneous single-body energy spectrum,
presents particle-hole symmetry \cite{Mula.22}, i.e., the eigenvalues
of $h_{ij}(t)$ fulfill that

\beq
\varepsilon_k(t)=-\varepsilon_{N+1-k}(t).
\eeq
Moreover, the information spread along the chain is limited by the
Lieb-Robinson velocity, which corresponds to the Fermi velocity in our
model, $v_F=1$ \cite{Lieb.72}.

\bigskip

Let us consider a block $A$ composed of the left-most $\ell$ sites in
the chain, $A=\{1,\cdots,\ell\}$, when the state is a Slater
determinant $\ket|\Psi>$, as in Eq. \eqref{eq:slater}. Its
entanglement properties are determined by the $N\times N$ correlation
submatrix, $C_A$, defined by

\beq
(C_A)_{ij}=\sum_{k=1}^{N/2} \bar U_{ki} U_{kj},
\eeq
with $i$, $j\in A$ and zero otherwise. We realize that this matrix can
be built as the product of three projectors,

\beq
C_A = P_A C P_A,
\label{eq:projectorproduct}
\eeq
where $P_A$ is the $N\times N$ matrix that projects on the $\ell$
sites of block $A$, and $C$ is the full correlator matrix, which
projects on the occupied modes. Let the spectrum of $C_A$ be denoted
by $\{\nu^A_k\}$, where each eigenvalue can be proved to lie in
$[0,1]$, as it corresponds to a truncated projector. It determines the
entanglement entropy of block $A$, defined as

\beq
S_A\equiv -\Tr(\rho_A \log\rho_A),
\eeq
where $\rho_A=\Tr_{\bar A} \ket|\Psi>\bra<\Psi|$, through the
following expression

\beq
S_A=\sum_{k=1}^m \HH(\nu^A_k),
\eeq
where

\beq
\HH(x)\equiv -x\log(x)-(1-x)\log(1-x).
\label{eq:H2x}
\eeq


\section{Long term behavior of the stirred Dirac vacuum}
\label{sec:num}

In this section we characterize the physical properties of the
long-term stirred Dirac vacuum through numerical analysis of different
observables: the absorbed energy, the mode occupations and the
statistical properties of the Floquet spectrum.

\subsection{Absorbed energy and vacuum friction}

Let us consider the time-evolution of the total energy of the system,
defined by the expectation value of the instanteneous Hamiltonian,

\beq
E(t)=\<\Psi(t)|H(t)|\Psi(t)\>.
\label{eq:energy}
\eeq
Notice that our system is isolated, and energy is pumped into it
without any relaxation mechanism. Thus, we expect the total energy to
grow, at least initially, which can be interpreted as a {\em vacuum
  friction}, provided by whatever forces make the obstacle move. Yet,
as we will see, this energy growth must saturate at a certain moment,
since our system is finite.

\begin{figure}
  \includegraphics[width=8cm]{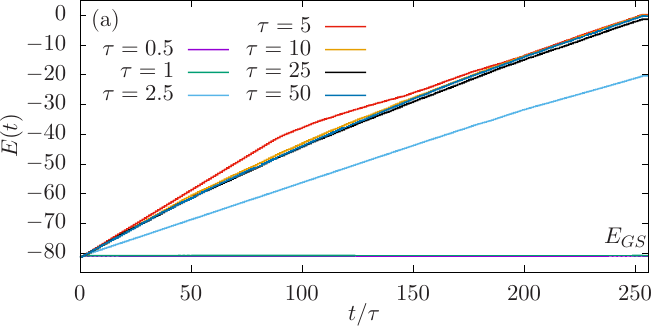}
  \vskip 2mm 
  \includegraphics[width=8cm]{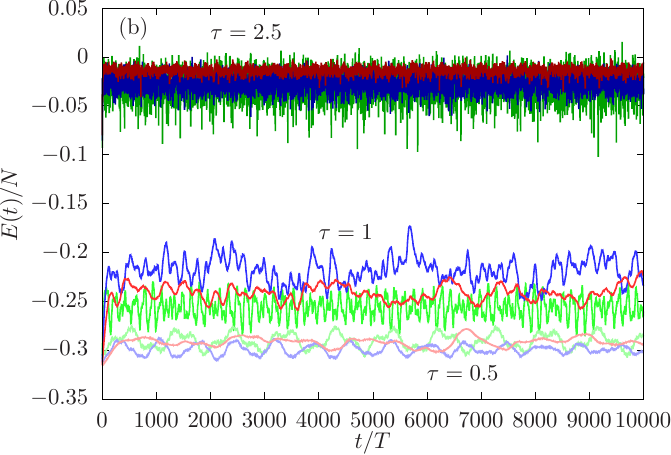}
  \caption{Expected value of total energy $\left<H(t)\right>$, Eq.
    \eqref{eq:energy}, as a function for time, for different values of
    the step-time $\tau$. (a) Short-time behavior within the first
    period, $t<T$, for $N=256$, as a function of $t/\tau$. Notice
    that, for fast schedules, $\tau=0.5$ and 1, the energy absorption
    is extremely low in comparison with the slow ones. The bottom
    energy level corresponds to the initial GS value, $E_{GS}$. (b)
    Long-time behavior as a function of $t/T$, showing the expected
    value of the energy per site at the beginning of each cycle for
    $N=64$ (green), $128$ (blue) and $256$ (red) and different color
    intensities for three values of $\tau=0.5$, 1 and 2.5, up to time
    $10^4\,T$. Notice that in all cases the energy fluctuations seem
    to be stationary.}
  \label{fig:energy}
\end{figure}

Fig. \ref{fig:energy} shows the numerical results for the expected
value of the energy $E(t)$ using $N=256$ for several step-times
$\tau$, as a function of $t/\tau$ so that they reach the first cycle
at the same point in the graph. The initial value can be estimated
analytically, $E(0)\approx -N/\pi$ to a good approximation for large
$N$ \cite{Mula.21}. Fig. \ref{fig:energy} (a) shows the short-time
absorption of energy, which is almost linear for short times.
Interestingly, for large values of $\tau$ the curves coincide, while
larger deviations can be found for intermediate values.

Fig. \ref{fig:energy} (b) displays the energy per site $E(t)/N$ at the
beginning of each cycle, for a much longer time-span, i.e. $10^4$ full
cycles, using three different sizes, $N=64$, 128 and 256, and three
values of $\tau=0.5$, 1 and 2.5. The plots suggest that a stationary
regime is reached for all the schedules, with an average value of the
energy that grows with $\tau$, saturating at $E\approx 0$. Moreover,
both the amplitude and the time scales associated to the fluctuations
depend both on $N$ and $\tau$. We may consider whether showing only
the energies at the starting point of each cycle is creating a bias.
Indeed, considering the full curve will add some fluctuations, but
their amplitude is always small, and we have not considered them in
the plots.

\begin{figure}
  \includegraphics[width=8cm]{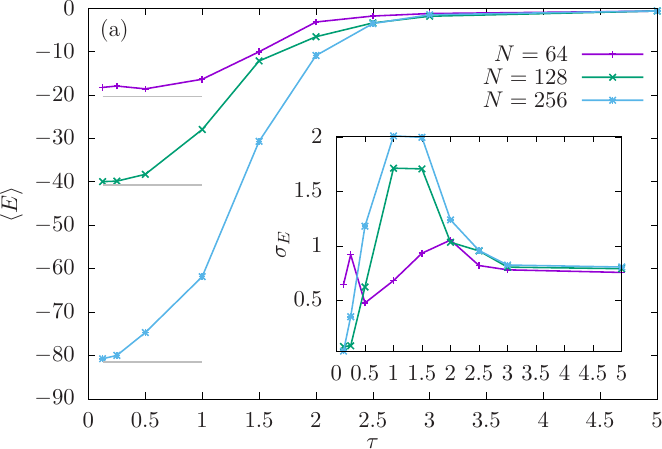}
   \includegraphics[width=8cm]{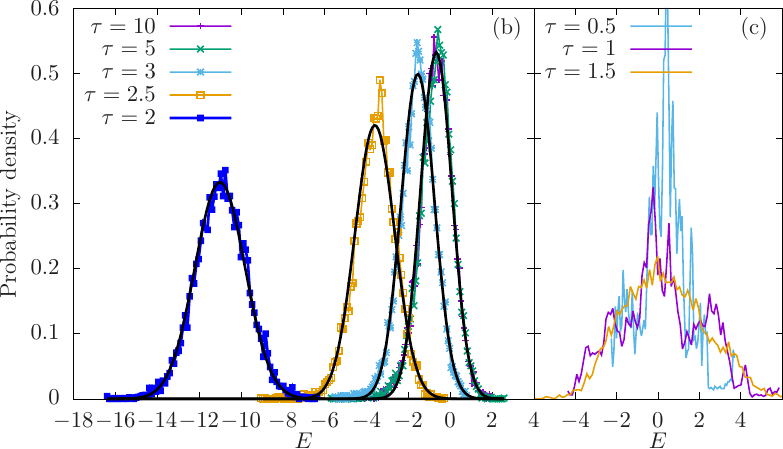}
   \caption{(a) Long-term expected value of the energy, $\<E\>$, as a
     function of the step-time $\tau$, for different system sizes,
     $N=64$, 128 and 256. The gray horizontal lines denote the initial
     values of the energy. The inset shows the standard deviation
     within the stationary regime. (b) Energy histograms for slow
     schedules, using always $N=256$, $\tau\geq 2$ (left) and fast
     schedules, $\tau\leq 1$ (right, shifted to have zero average),
     showing a Gaussian fit whenever suitable.}
  \label{fig:stats_E}
\end{figure}

Thus, we are naturally led to consider the energy values as a
stationary stochastic time-series for long enough times. Fig.
\ref{fig:stats_E} shows the statistical properties of the expected
value of the energy. Panel \ref{fig:stats_E} (a) shows the average
value of the energy within the stationary regime as a function of
$\tau$ for different system sizes $N$. We notice that, as $\tau$
grows, the expected value of the energy approaches zero, while for
$\tau\to 0$ the expected value approaches $-N/\pi$, which is the
initial energy of the system within a good approximation. The inset of
Fig. \ref{fig:stats_E} (a) show the standard deviation of the values
of the energy within the stationary regime, showing a maximum at
intermediate values of $\tau \approx 1$. 

It is worthy to examine the full histogram of the values taken by the
energy within the stationary regime, as they are shown in Fig.
\ref{fig:stats_E} (b). The left panel shows the histograms for
$\tau\geq 2$, i.e. the slow schedules, along with suitable Gaussian
fits. On the right panel we can see, superimposed, the converged
histograms for $\tau=0.5$, 1 and 1.5, shifted so that their average
becomes zero, which deviate substantially from the Gaussian
distribution as $\tau$ decreases.

\medskip

All these results lead us to conjecture that obstacle speeds larger
than the Lieb-Robinson velocity, i.e. $\tau<1$, lead to a fast regime
which differs substantially from the slow regime, characterized by
$\tau\gg 1$, in which (a) the expected value of the energy approaches
zero, (b) the energy fluctuations are Gaussian. Issue (a) can be
readily explained, considering that the system becomes effectively
thermalized at infinite temperature. Since the system respects
particle-hole symmetry for all times, the expected value of the energy
in this regime is zero, leading also naturally to Gaussian
fluctuations of the energy. On the other hand, non-equilibrium quantum
evolutions led by quantum coherences present typically non-Gaussian
fluctuations in the energy \cite{Chenu.18, Miller.19, Scandi.20,
  Miller.21, Zawadzki.23}.

\subsection{Structure in momentum space}

Let us consider $n_k=b^\dagger_k b_k$, i.e. the occupation operator
associated to the $k$-th mode of the clean Hamiltonian $H_0$, i.e.

\beq
b^\dagger_k = \sqrt{2\over N+1}
\sum_{i=1}^N \sin\({\pi i \over N+1}\) c^\dagger_i,
\eeq
such that $H_0=\sum_k \varepsilon_k n_k$, and $\varepsilon_k
=-2\cos(\pi k/(N+1))$. Initially, $\<n_k(0)\>\equiv
\<\Psi(0)|n_k|\Psi(0)\>=1$ if $\varepsilon_k <0$, and $0$ otherwise.
In other words, the mode occupations follow a step function,
$\<n_k\>=\theta(N/2-k)$.

\begin{figure}
  \includegraphics[width=8cm]{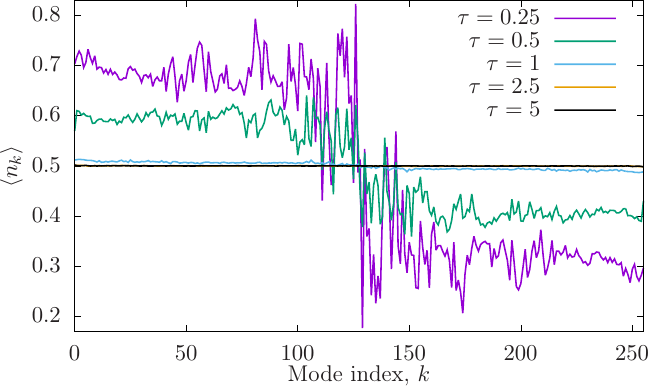}
  \caption{Long-term average occupation of the original Hamiltonian
    modes in the stationary regime, for different values of $\tau$,
    using always $N=256$.}
  \label{fig:occupation}
\end{figure}

We may wonder about the long-term evolution of $\<n_k(t)\>$. Thus, we have
obtained their time-averages within the stationary regime using
$N=256$ and different values of $\tau$, as it is shown in Fig.
\ref{fig:occupation}. Indeed, we can see that in the slow regime,
$\tau\gg 1$, the average occupations are flat, i.e. $\<n_k\> \approx
1/2$ for all $k$, while within the fast regime, $\tau<1$ we observe a
complex pattern which, notwithstanding, remains more and more similar
to the original step function as $\tau\to 0$.

\subsection{Floquet effective Hamiltonian}

The Hamiltonian imposed on our physical system is periodic,
$H(t+T)=H(t)$, and therefore it is relevant to ask about the effective
Floquet Hamiltonian, $H_F$, which is defined as the operator which
would provide the same evolution after a single period, and can
provide a lot of interesting information about the long-term behavior.

Let $U(t)$ be the evolution operator of our system, defined in
terms of a time-ordered exponential \cite{Eckardt.17,Tsuji.23}. Then,
we may define implicitly $H_F$ through

\beq
U(T)\equiv\exp(-i H_FT).
\eeq
The eigenvalues of $H_F$, $\{\epsilon_k\}$, are called {\em
  quasi-energies}, and are determined modulo $\Omega=2\pi/T$. The
properties of the Floquet Hamiltonian have been employed in order to
characterize quantum chaotic behavior \cite{Bohigas.84}, associating
Poisson level statistics to integrable systems and gaussian orthogonal
or unitary ensemble (GOE/GUE) statistics to chaotic ones
\cite{Berry.77,DAlessio.14,Scaramazza.16}.

\begin{figure}
  \includegraphics[width=8cm]{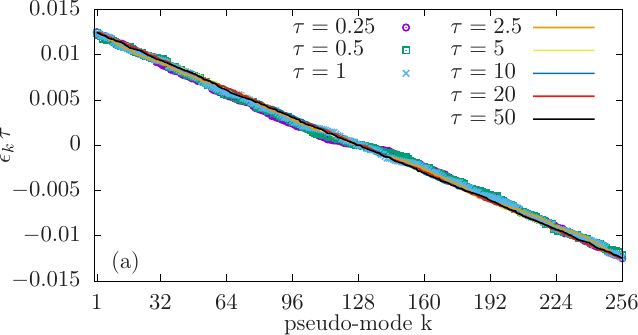}
  \vskip 2mm
  \includegraphics[width=8cm]{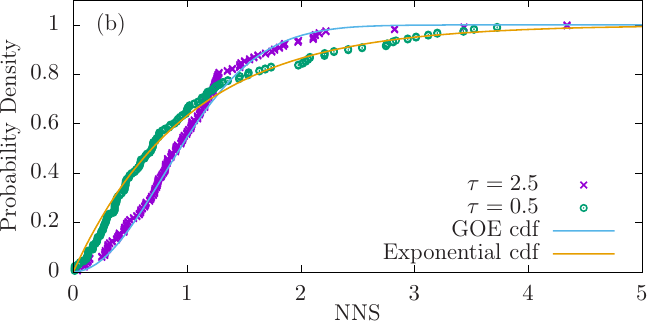}
  \vskip 2mm
  \includegraphics[width=8cm]{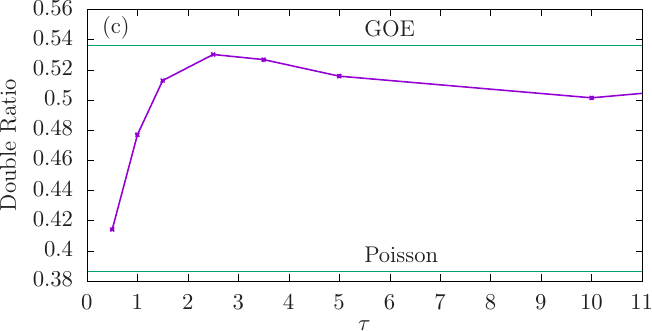}
  \caption{(a) Quasi-energy spectrum, $\tau\,\epsilon_k$ for different
    step-times $\tau$, using system size $N=256$ and time-steps
    ranging from $\tau=0.25$ up to $\tau=50$. The slow-driven regime
    is linear, with small non-linearities appearing around
    $\epsilon_k\approx 0$ for the fast-driven regime; (b) Cumulative
    distribution functions (CDF) of the level separations for
    $\tau=0.5$ and 2.5, along with the theoretically expected values
    corresponding to the Poisson and GOE distributions; (c) Double
    ratio of quasi-energy separations as a function of the step-time
    $\tau$ for $N=256$, along with the theoretical values
    corresponding to the same distributions.}
  \label{fig:spectrum}
\end{figure}

Fig. \ref{fig:spectrum} (a) shows this scaled quasi-energy spectrum,
$\epsilon_k \tau$ for different values of $N$ and $\tau$, showing an
approximate collapse. Indeed, the quasi-energy spectrum is
approximately linear throughout the range, presenting a slight
curvature for $\tau\ll 1$. Thus, it is therefore relevant to ask about
the level statistics. Fig. \ref{fig:spectrum} (b) shows the cumulative
distribution function (cdf) of the level separations, which are
defined as

\beq
s_k \equiv \epsilon_{k+1}-\epsilon_k,
\eeq
for $\tau=0.5$ and 2.5, using always $N=256$, along with the Poisson
distribution and the one associated to the gaussian orthogonal
ensemble (GOE) for comparison, which are respectively given by

\begin{itemize}
\item Poisson, $p(s)= e^{-s}$.
\item GOE, $p(s)=\frac{\pi}{2} s e^{\frac{\pi}{4}s^2}$.
\end{itemize}

In order to characterize the crossover between the two regimes we have
estimated the {\em double ratio} of the quasi-energies level
distribution, defined as the average of the ratios
\cite{Atas.13,Chavda.14}

\beq
\tilde{r}_n\equiv\frac{\min(s_n, s_{n-1})}{\max(s_n, s_{n-1})}=\min(r_n,1/r_n)
\eeq
where $r_n=\frac{s_{n+1}}{s_n}$ are the ratios of consecutive level
separations. The theoretical values for the average
$\left<\tilde{r}_n\right>$ for Poisson, GOE and GUE distributions are
as follows,

\begin{align}
  \left<\tilde{r}_n\right>_{\text{Poisson}} &\approx 0.38, \nn\\
  \left<\tilde{r}_n\right>_{\text{GOE}} &\approx 0.54, \nn\\
  \left<\tilde{r}_n\right>_{\text{GUE}} &\approx 0.59.
\end{align}

Fig. \ref{fig:spectrum} (c) shows the double ratio
$\left<\tilde{r}_n\right>$ for our model driven at different speeds
and using $N=256$, displaying a crossover very similar to the one
found in other situations more artificially constructed
\cite{Chavda.14}.

\subsection{Floquet occupations}
\label{ssec:floquet_occ}

Conserved quantities are of extreme importance when describing any
dynamical system. A time-independent free-fermion Hamiltonian always
commutes with the number operators for a series of modes, thus
providing us with $N$ conserved quantities, i.e. the {\em occupations}
for each mode. In our case, the Hamiltonian is time-dependent, but the
time evolution can be mimicked using the Floquet effective
Hamiltonian. We may thus define the Floquet occupations, $n^F_k\equiv
\<f^\dagger_k f_k\>$, where $H_F=\sum_k \epsilon_k f^\dagger_k f_k$ is
the Floquet Hamiltonian. If we only observe the system at times
$t_m=mT$, then these Floquet occupations are exactly preserved.

\begin{figure}
  \includegraphics[width=8cm]{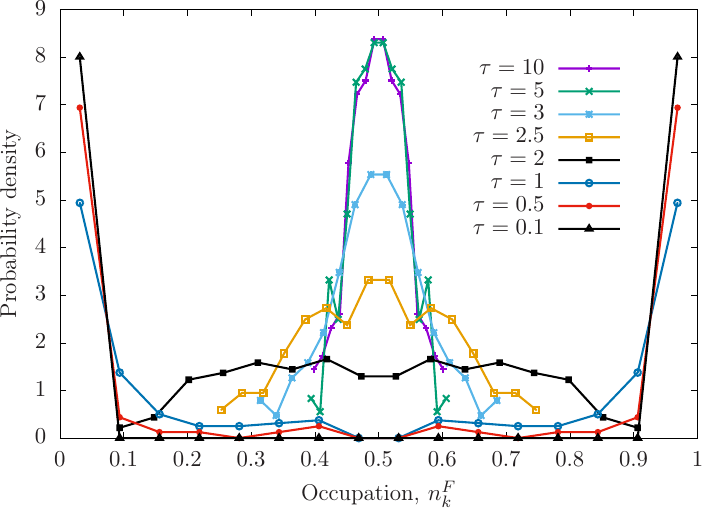}
  \caption{Histograms for the Floquet occupations, $n^F_k$, within the
    long-term evolution of a stirred Dirac chain with $N=256$ and
    different values of $\tau$. Notice a crossover between a fast
    regime, with a bimodal histogram concentrated on $n^F_k=0$ and
    $1$, and a slow regime, with the values concentrated on
    $n^F_k=1/2$.}
  \label{fig:floquet_occup}
\end{figure}

The values of the Floquet occupations, $\{n^F_k\}$, have been obtained
for the initial state, $\ket|\Psi(0)>$, using $N=256$ and several
values for $\tau$, and their histograms are shown in Fig.
\ref{fig:floquet_occup}. Interestingly, their behavior is very
different for slow and for fast schedules. When $\tau\ll 1$, the
histogram becomes bimodal, essentially concentrated near $n^F_k=0$ and
1. On the other hand, for very slow schedules, $\tau\gg 1$, the
histogram is concentrated around $n^F_k=1/2$, reaching a finite value
for the deviation in the large $\tau$ limit.

\section{The random Slater ensemble}
\label{sec:rmt}

Along the next sections we will argue that the stationary state for
the slow schedule, $\tau\gg 1$, can be described as a statistical
ensemble. Therefore, in this section we will present some basic facts
about random fermionic Gaussian states.

Choosing a suitable unitary-invariant measure, we can pick up a random
state in a Hilbert space of $N$ qubits and find the expected value of
its entanglement entropy when we separate a block of $\ell$ qubits. The
results is known as Page's law \cite{Page.93},

\beq
\<S(\ell)\> \approx \ell\log 2 - {1\over 2^{N-2\ell+1}}.
\label{eq:page}
\eeq
If our state is a Slater determinant, then Eq. \eqref{eq:page} does
not apply. Different researchers have considered this extension to the
Page problem \cite{Bianchi.21,Huang.22}, and we will provide in this
section a simple explanation of the main results.

\medskip

A Slater determinant is fully characterized by its correlation matrix,
$C_{ij}=\langle c^\dagger_i c_j \rangle$, which can be regarded as a
projector on a subspace of dimension $m$, the number of particles.
Concretely, its reduced density matrix associated to a block $A$ of
size $\ell$ can be obtained by considering the associated submatrix,
as it was expressed in Eq. \eqref{eq:projectorproduct}, $C_A=P_A C
P_A$, where $P_A$ is the projector on the sites which belong to $A$.
The spectrum of $C_A$, denoted by $\{\nu^A_k\}$, determines the
entanglement spectrum and the entanglement entropy of block $A$ in
that state.

Let us consider the ensemble of Slater determinants on $N$ sites with
$m$ fermionic particles, which are chosen according to the Haar
measure in U($N$), i.e. we choose a random unitary matrix $V$ from the
gaussian unitary ensemble (GUE), truncate its first $m$ columns, and
form the correlation matrix $C=VV^\dagger$. Alternatively, we may
write an $N\times m$ matrix with Gaussian entries (with zero mean and
unit variance both for the real and imaginary parts), and let it
undergo a Gram-Schmidt procedure. Any such submatrix $C_A$, of
dimension $\ell\times\ell$, is said to belong to the {\em Jacobi
  ensemble} \cite{Collins.04}. If we let $m=N/2$ and define
$\mu=\ell/N$, assuming that $\mu\leq 1/2$, we can obtain an expression
for the eigenvalue density, which simplifies slightly if we define
$\lambda^A_k\equiv 2\nu^A_k-1$. Indeed,

\beq
\rho_\mu(\lambda)={\sqrt{4\mu(1-\mu)-\lambda^2}\over
  2\pi\mu(1-\lambda^2)},
\label{eq:spectrum_dens}
\eeq
which is only defined in the interval between $\lambda_\pm
(\mu)\equiv\pm 2\sqrt{\mu(1-\mu)}$. If the eigenvalues were
uncorrelated, the average entanglement entropy would be written as

\beq
\<S_A\> \approx \mu N \<\HH(\mu)\>.
\eeq
with

\beq
\<\HH(\mu)\> = \int_{\lambda_-(\mu)}^{\lambda_+(\mu)}
\HH\({\lambda+1\over 2}\)\;
\rho_\mu(\lambda)\, d\lambda.
\label{eq:avh2}
\eeq
Appendix \ref{sec.Calculation of H2_Exact} proves that

\beq
\<\HH(\mu)\> = \log(2) - 1 - {(1-\mu)\over \mu}\log(1-\mu),
\label{eq:h2_exact}
\eeq
thus leading to

\beq
S(\ell) \approx \ell \log(2) - \ell - (N-\ell) \log\(1-{\ell \over N}\).
\label{eq:entropy_approx}
\eeq
For the half-chain, $\ell=N/2$, we have

\beq
S(N/2) \approx N \(\log(2)-{1\over 2}\),
\eeq
which is lower than the Page law equivalent. Indeed, the entropy per
site appears to be $2\log(2)-1\approx 0.386$ in the random Slater
ensemble while it is $\log(2) \approx 0.693$ in the Page ensemble.

\medskip

The results presented above are only approximate, because the
eigenvalues from a single matrix can not be considered to be
uncorrelated. The exact statitical properties for the entropy were
obtained in \cite{Bianchi.21,Huang.22} and present a small correction
with respect to our simple approximation. In our notation,

\begin{align}
\langle S(\ell) \rangle&=
1-\mu(1+N)-m\mu\Psi(m)+N\Psi(N)\nn\\
&+\mu(m-N)\Psi(N-m)+(\ell-N)\Psi(N-\ell+1)
\label{eq:entropy_exact}
\end{align}
for $\ell\leq N/2$, where $\Psi(x)=\Gamma'(x)/\Gamma(x)$ is the
digamma function. Its variance, in turn, can be expressed as

\begin{align}
  (\Delta S_A)^2&=
  \log(1-\mu)+\mu+\mu^2\nn\\
  &+\mu^2\(2\frac{m}{N}-1\)\log\(\frac{N}{m}-1\)\nn\\
  &+\mu(\mu-1)\(\frac{m}{N}-1\)\frac{m}{N}\log^2\(\frac{N}{m}-1\)+O(1),
\end{align}
which reduces to $(\Delta S_A)^2=\log(1-\mu)+\mu+\mu^2$ in the case of
half-filling.

\medskip

\begin{figure}
  \includegraphics[width=8cm]{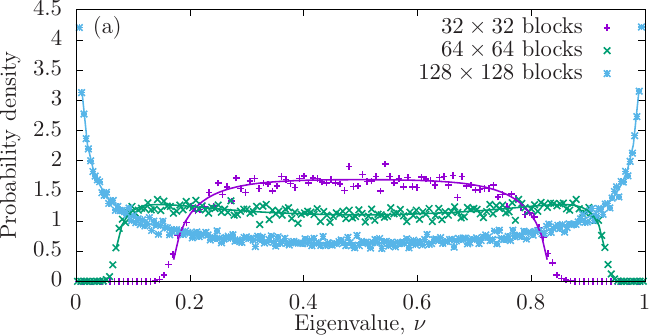}
  \vskip 2mm 
  \includegraphics[width=8cm]{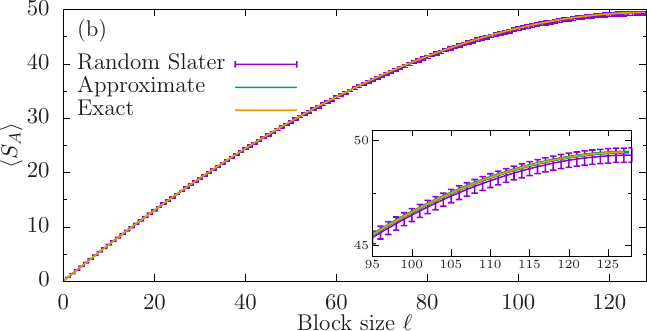}
  \caption{(a) Spectral density for random Slater truncated
    correlation matrices with $N=256$ and different values of the
    block size $\ell$, comparing with Eq. \eqref{eq:spectrum_dens};
    (b) Expected value of the entanglement entropy obtained from the
    same data, along with the two approximate and exact theoretical
    curves, Eq. \eqref{eq:entropy_approx} and
    \eqref{eq:entropy_exact}, with their errorbars. Inset: zoom up of
    the upper-right part of the panel, in order to highlight the small
    differences between the curves.}
  \label{fig:random}
\end{figure}

We have performed numerical experiments with random unitary matrices
in order to check the validity of Eqs. \eqref{eq:spectrum_dens},
\eqref{eq:entropy_approx} and \eqref{eq:entropy_exact}, and the
results are shown in Fig. \ref{fig:random}. Panel (a) shows the
eigenvalue histogram of 500 matrices sampled from systems with $N=256$
and different values of $\ell$. Panel (b) shows the expected value of
the entanglement entropy from our numerical experiments, comparing to
our approximation, Eq. \eqref{eq:entropy_approx}, and the exact
theoretical prediction, Eq. \eqref{eq:entropy_exact}. We may conclude
that our simple approximation is quite accurate.

\section{Entanglement in the stirred Dirac vacuum}
\label{sec:entang}

Let us now discuss the statistical properties of the entanglement
structure in the long-term stationary state of the stirred Dirac
vacuum considered in Sec. \ref{sec:num}. As we will show, the slow and
the fast schedules present very different behaviors, and the slow
phase corresponds to the random Slater ensemble discussed in the
previous section.

\medskip

\begin{figure}
  \includegraphics[width=8cm]{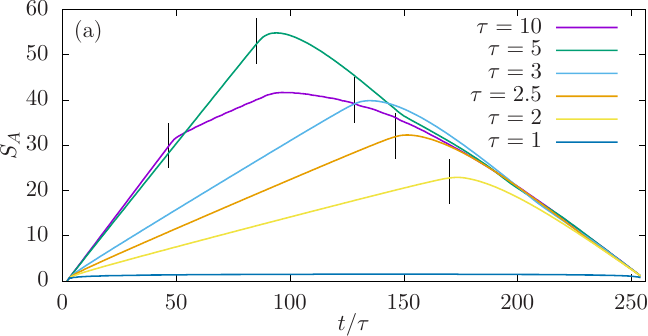}
  \vskip 2mm
  \includegraphics[width=8cm]{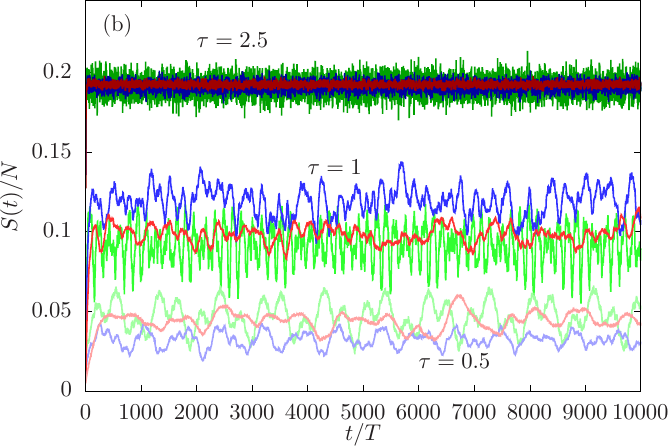}
  \caption{(a) EE of the partition associated to the obstacle for
    $N=256$, and a range of values of the step-time $\tau$. The
    vertical marks denote the expected time for the rebound of the
    quench information; (b) EE of the half chain as a function of
    time, within a larger time-span using $N=64$ (green), 128 (blue)
    and 256 (red), and using color intensity to denote different
    values of $\tau=0.5$, 1 and 2.5.}
  \label{fig:entropy}
\end{figure}

Fig. \ref{fig:entropy} (a) shows the time evolution of the EE
associated to the partition defined by the position of the obstacle,
as a function of time divided by $\tau$ (i.e. the position), for
different values of $\tau$, using always $N=256$, and only for the
first period, $t\leq T$. Notice that the initial growth rate increases
with the time-step. Yet, the maximum is reached at a time which
decreases with $\tau$, corresponding to the moment in which the
information of the initial quench ---which travels at the
Lieb-Robinson velocity--- reaches the opposite boundary, bounces back
and meets the obstacle again. Simple kinematic arguments show that
this time, $t_B$, fulfills

\beq
t_B = 2L{\tau \over \tau+1},
\eeq
which is signaled by the vertical marks in Fig. \ref{fig:entropy} (a).
The second knee in the entropy curve for $\tau=10$ corresponds to the
second rebound, but further rebounds are no longer coherent enough to
show up in the curve.

Alternatively, we can also consider the time evolution of the EE per
site of a fixed partition, e.g. the left half, as it is shown in Fig.
\ref{fig:entropy} (b), on a larger time scale, as a function of $t/T$.
It can be observed that, in the long run, the entropy also reaches a
stationary regime, as we found for the energy, and also in this case
the long-term average and deviation depends both on $N$ and $\tau$.

\begin{figure}
  \includegraphics[width=8cm]{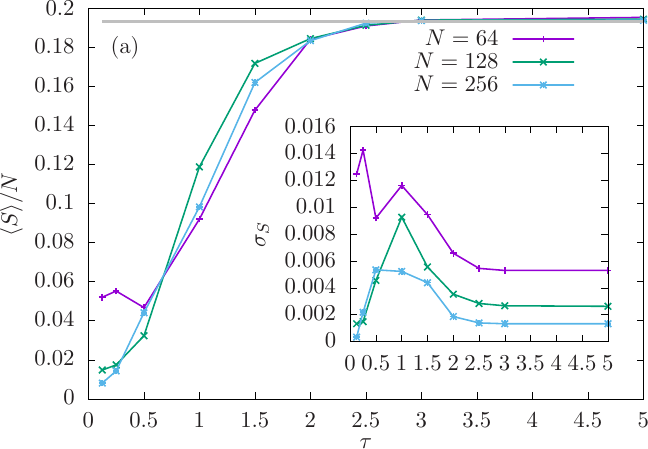}
   \includegraphics[width=8cm]{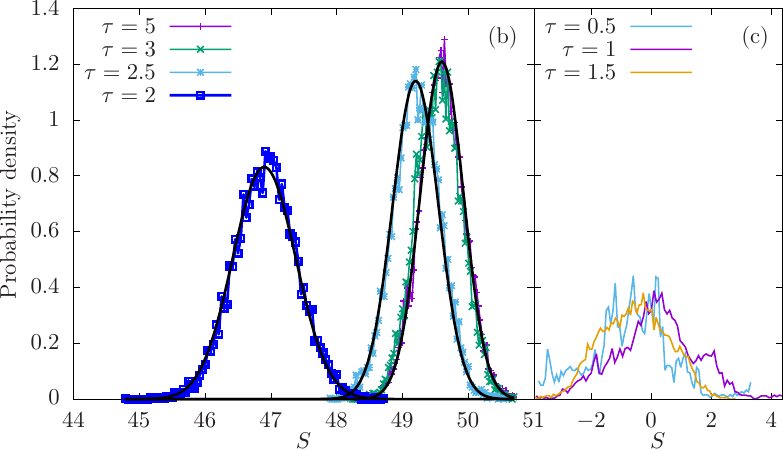}
   \caption{(a) Long-term expectaction value of the half-chain entropy
     per site, $\<S\>$, as a function of the step-time $\tau$, for
     different system sizes, $N=64$, 128 and 256. The gray horizontal
     lines denote the theoretical prediction for the long-term entropy
     per site. The inset shows the standard deviation within the
     stationary regime. (b) Entropy histograms for slow schedules,
     using always $N=256$, $\tau\geq 2$ (left) and fast schedules,
     $\tau\leq 1$ (right), showing a Gaussian fit whenever suitable.}
  \label{fig:stats_S}
\end{figure}

Following the analysis performed for the entropy, we plot in Fig.
\ref{fig:stats_S} (a) the long-term average of the half-chain entropy
per site as a function of $\tau$ for $N=64$, 128 and 256. We again
observe a crossover between a low entropy phase for fast schedules,
$\tau<1$ and a high entropy phase for slow schedules, $\tau\gg 1$. The
theoretically expected value of the entropy per site, $\log(2)-1/2$,
is shown as a horizontal grey line. 

Fig. \ref{fig:stats_S} (b) shows the entropy histogram for $N=256$ and
larger values of $\tau$ on its left panel, $\tau=2$, 2.5, 3 and 5,
along with a suitable Gaussian fit. On the right panel we observe the
histogram of the entropy for lower values of the time-step $\tau$,
showing that they do not present a Gaussian shape.

\medskip

\begin{figure}
  \includegraphics[width=8cm]{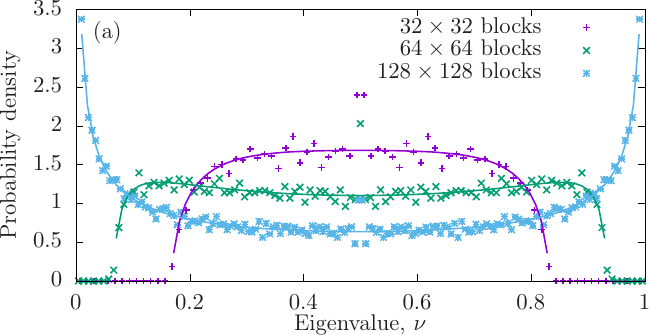}
  \vskip 2mm
  \includegraphics[width=8cm]{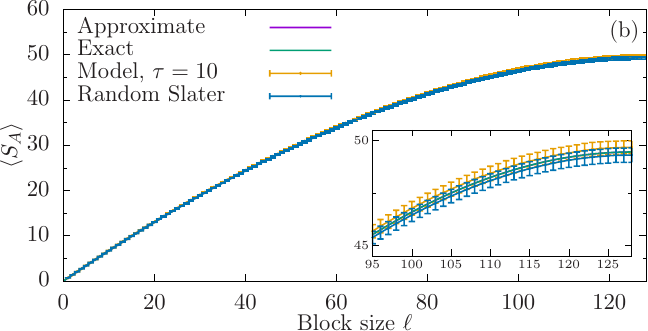}
  \caption{(a) Spectral density for the stirred Dirac vacuum for the
    truncated correlation matrices with $N=256$ and different values
    of block size $\ell$, comparing with Eq. \eqref{eq:spectrum_dens},
    for a large step-time $\tau=10$; (b) Expected value of the
    entanglement entropy obtained from the same data, along with the
    two approximate and exact theoretical curves, Eq.
    \eqref{eq:entropy_approx} and \eqref{eq:entropy_exact}. The inset
    highlights the tiny difference between the theoretical
    calculations, and between those and the numerical results for the
    random Slater model and the stirred Dirac vacuum.}
  \label{fig:EE_toy}
\end{figure}

Now, let us address our main question in this section: does the
stationary regime of the stirred Dirac vacuum correspond to the Jacobi
class? The best approach is to compare the predictions for the EE and
for the eigenvalues of the truncated correlation matrices, as we do in
Fig. \ref{fig:EE_toy}. Indeed, in panel (a) we observe that the
average values of the spectral density for the aforementioned
correlation matrices follow the Jacobi law, Eq.
\eqref{eq:spectrum_dens}, using $N=256$, $\tau=10$ and three block
sizes. Moreover, the average values of the EE also follow the
predictions of Eq. \eqref{eq:entropy_approx} and
\eqref{eq:entropy_exact}. In this case, we plot both theoretical
curves along with the averages obtained for the random Slater ensemble
and the stirred Dirac vacuum using again $N=256$ and $\tau=10$. The
inset highlights the tiny differences which can be observed between
them.

\medskip

\begin{figure}
  \includegraphics[width=8cm]{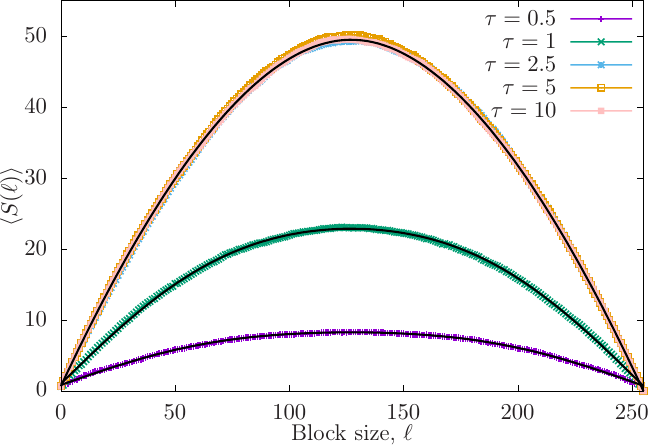}
  \caption{Long-term average of the entropy profiles for $N=256$ and
    different values of $\tau$, along with the theoretical
    expectations. For $\tau\gg 1$ we have shown Eq.
    \eqref{eq:entropy_approx} in black, while for lower values of
    $\tau$ we have fit to the form \eqref{eq:3sine}.}
  \label{fig:profile_S}
\end{figure}

Let us also consider the average entropy profile in the stationary
state for different values of $\tau$, which is shown in Fig.
\ref{fig:profile_S}, which plots the value of $\<S(\ell)\>$ for
$N=256$. We can see that for $\tau=2.5$, 5 and 10 the entropy fits the
theoretical prediction (shown in black). For $\tau=1$ and $\tau=0.5$
the entropy is much lower, and we have fitted it to a different law
\cite{Fagotti.11},

\beq
S(\ell) \approx A + B\sin\( {\pi \ell\over N} \) + C \sin\(
{3\pi\ell\over N}\).
\label{eq:3sine}
\eeq

\medskip

\begin{figure}
  \includegraphics[width=8cm]{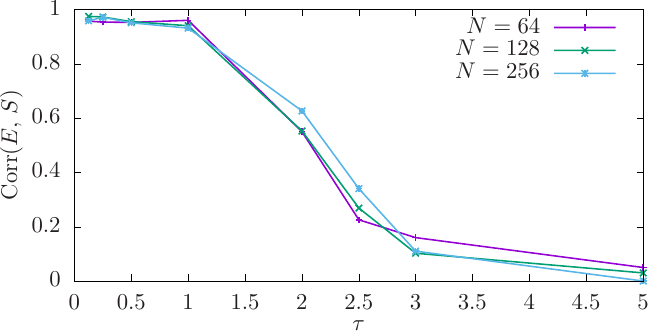}
  \caption{Correlation coefficient of entropy and energy for the
    random Slater model, as a function of $\tau$ for different values
    of $N$.}
  \label{fig:corrES}
\end{figure}

Finally, we have considered the joint statistical properties of the
entropy and the energy, and measured the correlation coefficient of
the corresponding time series, defined as

\beq
\text{Corr}(E,S)\equiv {\<ES\>\over \sigma(E)\sigma(S)}.
\eeq
The results are shown in Fig. \ref{fig:corrES}, as a function
of $\tau$ for different system sizes. We see that in all cases the
correlation is close to one for fast schedules, and close to zero for
$\tau\gg 1$, thus putting forward another feature of the slow schedule
phase, the independence of the fluctuations of $S$ and $E$.

\bigskip

A relevant question is in order. Since the system possesses $N$
exactly preserved quantities, i.e. the Floquet occupations shown in
Sec. \ref{ssec:floquet_occ}, how is it possible that the long-term
state of the stirred Dirac vacuum resembles the RSE for large $\tau$?
The answer is that, in the slow regime, the values of the Floquet
occupations {\em resemble} the expected values within the RSE, which
is always $1/2$. Thus, we may conjecture that the actual ensemble
explored in the long-term by the stirred Dirac vacuum can be described
by a generalized RSE, in similarity to the generalized Gibbs ensemble,
in which we force the expected values of a certain set of occupations
to take values far from 1/2.


\section{Entanglement links and the geometry blurring transition}
\label{sec:links}

The nature of the transition between the fast and slow schedules can
be very clearly characterized using the {\em entanglement link} (EL)
representation, which was recently introduced by some of us
\cite{Singha.20,Singha.21,Santalla.23}. The key insight behind the EL
is to take seriously the area law of entanglement, and to propose the
existence of an adjacency matrix $J_{ij}$ which approximately
represents the entanglement entropy of every block $A$,

\beq
S_A \approx \sum_{i\in A, j\in \bar A} J_{ij}.
\eeq
The EL representation is only exact in a few cases, such as valence
bond states \cite{Singha.20}, but it is surprisingly accurate in most
situations, including typical ground states and time-dependent states
after a quantum quench \cite{Santalla.23}. Even random states possess
reasonably accurate EL representations \cite{Singha.21}. The EL of the
GS of a gapped chain can be proved to be exponentially concentrated
along the main diagonal, while for a critical homogeneous chain the EL
fall as a power law, $J_{ij}\sim |i-j|^{-2}$, still showing clear
signals of the original geometry of the Hamiltonian. The EL can be
numerically obtained for Slater determinants in a fast way
\cite{Singha.21}. Defining $S_{i,j}$ as the entanglement entropy of
the block $\{i,i+1,\cdots,j-1\}$ (with periodic boundaries), we obtain

\beq
J_{ij} = S_{i,j} - S_{i+1,j} - S_{i,j+1} + S_{i+1,j+1}.
\eeq

\begin{figure*}
  \includegraphics[width=15cm]{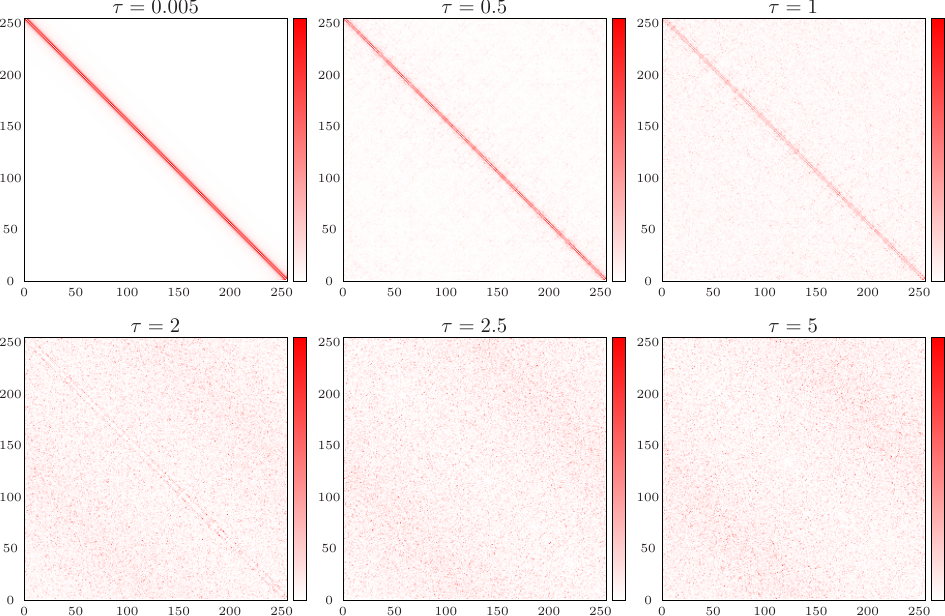}
  \caption{EL matrices corresponding to a chain with $N=256$ and
    different values of $\tau=0.005$, 0.5, 1, 2, 2.5 and 5, for
    $t=1275\,T$. The colorbox has always the same range, saturating at
    an EL intensity of 0.05.}
  \label{fig:links}
\end{figure*}

In Fig. \ref{fig:links} we show the EL matrix for $t=1275\,T$ using
$N=256$ and several values of $\tau$. We can see that in the fast
cases, i.e., $\tau=0.005$, 0.5 and 1, the EL matrix preserves the 1D
structure that the state inherits from the original Hamiltonian. Yet,
in the slow case the entanglement links appear to be homogeneously
spread, thus showing that the original geometry has been effectively
blurred.

\begin{figure}
  \includegraphics[width=8cm]{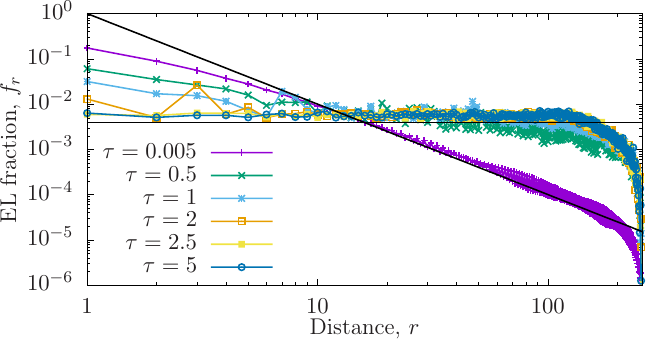}
  \caption{Fraction of the EL along the different subdiagonals for
    $N=256$ and $t=1275\,T$, for different values of $\tau$. The
    straight lines denote the expected power-law behavior for low
    $\tau$, $r^{-2}$, and the constant behavior $f_r\sim 1/N$ for
    large values of $\tau$.}
  \label{fig:links_transition}
\end{figure}

The geometry blurring transition may be further characterized by the
fraction of entanglement links along the different subdiagonals, i.e.

\beq
f_r\equiv{\sum_{i=1}^{N-r} J_{i,i+r} \over \sum_{i,j} J_{i,j}},
\eeq
which should decay as $r^{-2}$ for fast schedules, and remain constant
($\sim 1/N$) for slow ones. Fig. \ref{fig:links_transition} shows
that, indeed, the values of $f_r$ become constant for large values of
$\tau$, except for the final lattice effects, hinting at the idea that
the geometry has been blurred.

\section{Conclusions and further work}
\label{sec:conclusions}

We have considered the long-term behavior of the (1+1)D stirred Dirac
vacuum, which is defined as a free-fermionic chain in its ground
state, with an obstacle traveling through it, spending a time $\tau$
on each link. In classical terms, the motion of the obstacle would be
completely free, but in quantum terms we observe a drag force which
transfers energy to the state. After some time, a stationary regime is
reached, which we have characterized using different tools. In all
cases, we observe a different behavior in the {\em slow} and in the
{\em fast} regimes, defined by comparing the obstacle velocity and the
Lieb-Robinson velocity of the system.

In the slow regime, the average energy reaches the value zero, which
corresponds to a thermal state at infinite temperature. Also, the
Floquet effective Hamiltonian presents a GOE spectrum, which suggests
a connection to quantum chaos. Furthermore, the state can be
accurately represented by the {\em random Slater ensemble}, which has
been recently described, in which the occupied orbitals are randomly
chosen according to a Haar measure. The entanglement spectrum
histogram and the average entanglement entropy can be predicted using
random matrix theory, making use of the Jacobi ensemble. We have shown
that, indeed, the entanglement entropy slow schedules corresponds to
the random Slater prediction. 

It is interesting to notice that, in all cases, the system possesses a
large number of exactly conserved quantities, corresponding to the
occupations of the Floquet modes. Yet, only in the slow regime, the
expected values of these observables roughly coincide with their
expected values within the random Slater ensemble. We conjecture that
the long-term evolution can be described in both regimes as a
generalized random Slater ensemble, in which the expected values of
these occupations are forced towards their values on the initial
state. 

The physical picture can be ascertained by looking at the {\em
  entanglement links} (EL), a recently introduced tool which allows us
to represent the EE of all possible blocks from a weighted adjacency
matrix, to a good approximation. Indeed, the EL matrix allows the
geometry associated to the entanglement to become manifest. For the
fast regime, the EL structure is still one-dimensional, but for the
slow phase the EL spread, and the entanglement geometry blurs. The
system, effectively, {\em forgets} that it was 1D.

This work opens up several questions. The full characterization of the
fast regime should be performed extending the random Slater ensemble
in a suitable way through the use of the conserved quantities.
Moreover, integrability plays an important role in our physical
system, so it is relevant to consider what happens in its absence.
Also, it is interesting to consider a continuous movement of the
obstacle, instead of discrete, or to attenuate the effect of the
obstacle, making the hopping at the affected link reduced to a certain
value, instead of dropping to zero.

Also, we intend to consider different types of motion of the obstacle,
other than moving it with a constant speed, and to apply this
formalism to the case of the vacuum on a curved background metric,
which in many cases just amounts to an inhomogeneous set of hopping
parameters \cite{Boada.11,Laguna.17,Mula.21}.


\begin{acknowledgments}
We would like to acknowledge G. Sierra and N. Samos for very useful
discussions. This work was funded by the Spanish government through
grants PGC2018-094763-B-I00, PID2019-105182GB-I00 and
PID2021-123969NB-I00. 
\end{acknowledgments}


\onecolumngrid
\appendix\section{Explicit calculation of Eq. \eqref{eq:h2_exact}}
\label{sec.Calculation of H2_Exact}

We rewrite \eqref{eq:avh2} using \eqref{eq:H2x}, and we obtain
\begin{equation}
\<\HH(\mu)\> 
=
-\frac{1}{4\pi\mu}
\int_{-a}^a
\left[
\frac{\log\left(\frac{1-\lambda}{2}\right)}{1+\lambda}
+
\frac{\log\left(\frac{1+\lambda}{2}\right)}{1-\lambda}
\right]
\sqrt{a^2-\lambda^2}\,d\lambda
=
-\frac{1}{2\pi\mu}
\int_{-a}^a
\frac{\log\left(\frac{1+\lambda}{2}\right)}{1-\lambda}
\sqrt{a^2-\lambda^2}\,d\lambda,
\end{equation}
where we use the variable  $a=2\sqrt{\mu(1-\mu)}$ for convenience, and
we have made the change of variable $\lambda\mapsto-\lambda$ in the
first term. Since $0<\mu<\frac{1}{2}$, we have $0<a<1$.  To evaluate
this integral, we consider the complex valued function

\[
f_1(z)=\frac{\log\(\frac{1+z}{2}\)}{1-z}(z^2-a^2)^{1/2},
\qquad z\in\mathbb{C},
\]
where $0<a<1$, the root has a branch cut on $[-a,a]$, and the
logarithm has a branch cut on $(-\infty,-1]$.

\begin{figure}[h]
 	\includegraphics[width=7cm]{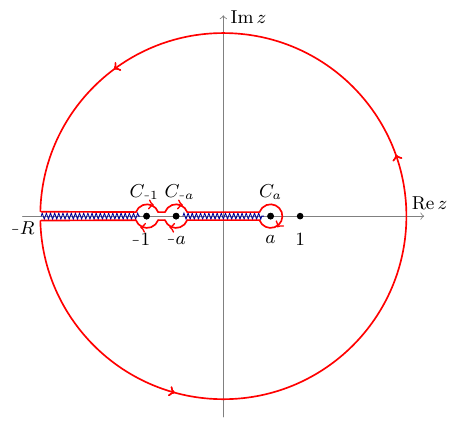}
 	\caption{Contour for complex integration.}
 	\label{fig:Contour_I1}
\end{figure}

We fix $R>a$ and we take an indented contour along the real axis,
avoiding the branch points $z=\pm a$ and the singularity at $z=-1$
with small circles of radius $\varepsilon$, and winding around the
pole at $z=1$ once in counterclockwise direction with a circular arc
$C_R$ of radius $R$ centered at the origin. See Figure
\ref{fig:Contour_I1}.

The boundary values of the function $f_1(z)$ (positive on the left and
negative on the right of the contour, with the given orientation) are
\[
\begin{aligned}
f_{1\pm}(x)
=
\begin{cases}
\displaystyle -\frac{\log\left(-\frac{1+x}{2}\right)\pm\pi i}{1-x}\, \sqrt{x^2-a^2},& \qquad  x\in (-\infty,-1-\varepsilon),\\[2mm]
\displaystyle -\frac{\log\left(\frac{1+x}{2}\right)}{1-x}\, \sqrt{x^2-a^2},& \qquad  x\in (-1+\varepsilon,-a-\varepsilon),\\[2mm]
\displaystyle \pm i \frac{\log\left(\frac{1+x}{2}\right)}{1-x}\, \sqrt{a^2-x^2},& \qquad  x\in (-a+\varepsilon,a-\varepsilon).
\end{cases}
\end{aligned}
\]

The integrals on the small circles tend to $0$ as $\varepsilon\to 0$,
writing $z=\pm a+\varepsilon e^{i\theta}$ or $z=-1+\varepsilon
e^{i\theta}$, and the function $f_1(z)$ has a simple pole at $z=1$.
The residue theorem implies that
\begin{equation}\label{eq:residuegeneralf11}
2i \int_{-a}^a  \frac{\log\left(\frac{1+x}{2}\right)}{1-x}\, \sqrt{a^2-x^2}dx
-2\pi i \int_{-R}^{-1}  \frac{\sqrt{x^2-a^2}}{1-x} dx
+
\int_{C_R} f_1(z)dz 
=
2\pi i\,\textrm{res}_{z=1} f_1(z)
=
2\pi i\,  \lim_{z\to 1} (z-1)f_1(z)
=
0, 
\end{equation}
and as a consequence

\begin{equation}\label{eq:residuegeneralf12}
\int_{-a}^a \frac{\log\left(\frac{1+x}{2}\right)}{1-x}\, \sqrt{a^2-x^2}dx
=\pi \int_1^R  \frac{\sqrt{x^2-a^2}}{1+x}dx
-\frac{1}{2i}\int_{C_R} f_1(z)dz.
\end{equation}
In order to evaluate the right hand side, we want to take the limit
$R\to\infty$, but both the integral on $(1,R)$ and the integral on
$C_R$ diverge, so we subtract and add the two leading terms:

\[
\begin{aligned}
\pi\int_1^R  \frac{\sqrt{x^2-a^2}}{1+x}dx
&=
\pi\int_1^R  \left[\frac{\sqrt{x^2-a^2}}{1+x}-1+\frac{1}{x}\right]dx
+\pi\int_1^R \left[1-\frac{1}{x}\right]dx\\
&=
\pi\int_1^R  \left[\frac{\sqrt{x^2-a^2}}{1+x}-1+\frac{1}{x}\right]dx
+\pi R-\pi-\pi\log R,
\end{aligned}
\]
and now the integral is convergent on $(1,R)$. Similarly, 

\[
\begin{aligned}
-\frac{1}{2i}
\int_{C_R} f_1(z)dz
&=
-\frac{1}{2i}
\int_{C_R}  \frac{\log\left(\frac{1+z}{2}\right)}{1-z} (z^2-a^2)^{1/2}dz\\
&=
-\frac{1}{2i}
\int_{C_R} \left[\frac{\log\left(\frac{1+z}{2}\right)}{1-z} (z^2-a^2)^{1/2}+\log z-\log 2
+
\frac{\log z-\log 2+1}{z}\right]dz\\
&+
\frac{1}{2i}
\int_{C_R} \left[\log z-\log 2+\frac{\log z-\log 2+1}{z}\right]dz\\
&=
-\frac{1}{2i}
\int_{C_R} \left[\frac{\log\left(\frac{1+z}{2}\right)}{1-z} (z^2-a^2)^{1/2}+\log z-\log 2+\frac{\log z-\log 2+1}{z}\right]dz\\
&-\pi R+\pi \log R+\pi-\pi \log 2,
\end{aligned}
\]
writing $z=Re^{i\theta}$,in order to calculate the last integral. Thus, \eqref{eq:residuegeneralf12} becomes
\begin{multline}\label{eq:residuegeneralf13}
\int_{-a}^a \frac{\log\left(\frac{1+x}{2}\right)}{1-x}\, \sqrt{a^2-x^2}dx
=
\pi\int_1^R  \left[\frac{\sqrt{x^2-a^2}}{1+x}-1+\frac{1}{x}\right]dx\\
-\frac{1}{2i}
\int_{C_R} \left[\frac{\log\left(\frac{1+z}{2}\right)}{1-z} (z^2-a^2)^{1/2}+\log z-\log 2+\frac{\log z-\log 2+1}{z}\right]dz-\pi \log 2.
\end{multline}
Now the integrand on $C_R$ is $\mathcal{O}\left(\frac{\log
  z}{z^2}\right)$ as $z\to\infty$, so the integral over $C_R$ tends to
$0$ when $R\to\infty$. This leads to

\begin{equation}\label{eq:residuegeneralf14}
\int_{-a}^a \frac{\log\left(\frac{1+x}{2}\right)}{1-x}\, \sqrt{a^2-x^2}dx
=
\pi\int_1^{\infty}  \left[\frac{\sqrt{x^2-a^2}}{1+x}-1+\frac{1}{x}\right]dx
-\pi \log 2,
\end{equation}
which is a somewhat simpler integral, since $a$ (or $\mu$) does not
appear in the limits of integration. We can calculate the indefinite
integral

\begin{equation}\label{eq:intrealx}
J(x)
=\int \frac{\sqrt{x^2-a^2}}{1+x} dx
\end{equation}
directly: if we make the change of variable $x=a\cosh t$, we have

\begin{equation}\label{eq:intrealt}
J(t)=a^2
\int \frac{\sinh^2 t}{1+a\cosh t}dt.
\end{equation}

Now we write the hyperbolic functions in terms of exponentials, and
make the change $v=e^t$: 

\begin{equation}\label{eq:intrealv}
J(v)=
\frac{a}{2}\int  \frac{v^4-2v^2+1}{v^2(v^2+2v/a+1)}dv
=\frac{a}{2}\int  \left(1-\frac{\frac{2}{a}v^3+v^2}{v^2(v^2+2v/a+1)}\right)dv.
\end{equation}

The roots of the denominator are $v=0$ and 
\[
v_{\pm}=\frac{-1\pm\sqrt{1-a^2}}{a}.
\]
If we do partial fractions, we obtain
\[
\frac{\frac{2}{a}v^3+v^2}{v^2(v^2+2v/a+1)}
=
\frac{\frac{2}{a}}{v}-\frac{1}{v^2}-\frac{2\sqrt{1-a^2}}{a (v-v_+)}+\frac{2\sqrt{1-a^2}}{a (v-v_-)},
\]
and therefore
\[
J(v)=\frac{av}{2}-\log v-\frac{2}{av}-\sqrt{1-a^2}\log\left(\frac{v-v_-}{v-v_+}\right)+C.
\]
We recall that $x=a\cosh t$, so $2x=ae^t+ae^{-t}$. It follows that
$e^t=x\pm \sqrt{x^2-a^2}$, and we take the plus sign because
$x=\infty$ corresponds to $t=\infty$. Therefore,
$v=e^t=x+\sqrt{x^2-a^2}$.  

We combine the previous primitive with those corresponding to the terms $-1+1/x$ that fix the divergence at infinity, and then
\[
\lim_{x\to\infty}J(v)
=
\lim_{x\to\infty}
\left[\frac{av}{2}-\log v-\frac{2}{av}-\sqrt{1-a^2}\log\left(\frac{v-v_-}{v-v_+}\right)-x+\log x\right]
=
-\log\frac{2}{a}.
\]
At $x=1$, we have $v(1)=1+\sqrt{1-a^2}$, so we obtain
\[
J(1)=-1+\sqrt{1-a^2}-(1+\sqrt{1-a^2})\log(1+\sqrt{1-a^2})+\log a
\]
Therefore, we obtain 
\[
\begin{aligned}
\int_1^{\infty}  \left[\frac{\sqrt{x^2-a^2}}{1+x}-1+\frac{1}{x}\right]dx
&=
-\log\frac{2}{a}
+1-\sqrt{1-a^2}+(1+\sqrt{1-a^2})\log(1+\sqrt{1-a^2})-\log a\\
&=
-\log 2+1-(1-2\mu)+2(1-\mu)\log(2(1-\mu))\\
&=
\log 2-2\mu\log 2+2\mu+2(1-\mu)\log(1-\mu),
\end{aligned}
\]
using that $a=2\sqrt{\mu(1-\mu)}$, so $1-a^2=(1-2\mu)^2$. 
Replacing this into \eqref{eq:residuegeneralf14}, we obtain
\begin{equation}
\begin{aligned}
\<\HH(\mu)\> 
=
-\frac{1}{2\mu}\int_1^{\infty}  \left[\frac{\sqrt{x^2-a^2}}{1+x}-1+\frac{1}{x}\right]dx
+\frac{\log 2}{2\mu}
&=
-\frac{1}{2\mu}\left[\log 2-2\mu\log 2+2\mu+2(1-\mu)\log(1-\mu)\right]
+\frac{\log 2}{2\mu}\\
&=
\log 2-1-\frac{1-\mu}{\mu}\log(1-\mu),
\end{aligned}
\end{equation}
which proves the result.

\twocolumngrid

\end{document}